\documentclass[sn-mathphys,Numbered]{sn-jnl}

\usepackage{graphicx}%
\usepackage{multirow}%
\usepackage{amsmath,amssymb,amsfonts}%
\usepackage{amsthm}%
\usepackage{mathrsfs}%
\usepackage[title]{appendix}%
\usepackage{xcolor}%
\usepackage{textcomp}%
\usepackage{manyfoot}%
\usepackage{booktabs}%
\usepackage{algorithm}%
\usepackage{algorithmicx}%
\usepackage{algpseudocode}%
\usepackage{listings}%

\theoremstyle{thmstyleone}%
\theoremstyle{thmstyletwo}%

\theoremstyle{thmstylethree}%

\begin{document}

\title[Article Title]{Hidden hierarchy in the rheology of dense suspensions}


\author[1]{\fnm{Abhinendra} \sur{Singh}}\email{abhinendra.singh@case.edu}
\affil[1]{\orgdiv{Department of Macromolecular Science and Engineering}, \orgname{Case Western Reserve University}, \orgaddress{ \city{Cleveland}, \postcode{10040}, \state{OH}, \country{USA}}}


\abstract{Dense suspensions of fine particles are significant in numerous biological, industrial, and natural phenomena. They also provide an ideal tool to develop statistical mechanics description for out-of-equilibrium systems. Predicting the bulk response of such materials has been challenging since these systems often undergo liquid-solid transitions upon a small change in solid concentration or applied loading. Developing an understanding of the mechanisms that drive these phenomena has over the last several years led to a surge in research activity at the intersection of fluid mechanics, granular materials, driven disordered systems, tribology, and soft condensed matter physics. 
One central aspect that emerged is that these phenomena are due to a shear-activated or deactivated network of contacts between particles.
The perspective briefly presents the current state of understanding and challenges associated with relating the flow of material at the bulk scale with the microscopic physics at the particle scale.
}




\maketitle

\section{Introduction}\label{sec1}
Suspensions of fine particles in a liquid are present in many industrial, geotechnical, and biological phenomena with examples ranging from the transport of concrete, mudflow, and red blood cells~\cite{Denn_2018, Denn_2014, guazzelli_2018, Jerolmack_2019}.
In these settings, the solid fine particles are often found to be roughly equal proportion by volume, termed ``dense suspensions''~\cite{Morris_2020}.
Although thoroughly studied in many fields (Chemical Engineering, Material Science, Physics, etc.) and practically useful, a unifying constitutive model relating material properties, composition, and bulk response remains elusive; there is no analog of Navier-Stokes equation or Statistical Mechanics description of sheared dense suspensions.

Historically, the suspensions have been treated as a fluid mechanics problem, i.e., the dynamics have been dominated by the viscous stresses induced from fluid due to the relative motion of particles~\cite{Brady_1985, Wagner_2009}.
However, the experimentally observed features could not be explained using the fluid mechanics approach, particularly the reversible liquid-solid transition often demonstrated by cornstarch (c.f. Dr. Seuss's Oobleck).

The shear-induced reversible transformation of cornstarch or cement from a low-viscosity easily flowing liquid to much-enhanced viscosity (close to solid) is highly undesirable in  most cases, such as creating problems in coating and mixing, clogging of transport materials~\cite{Barnes_1989}.
Recently, this phenomenon has been leveraged in various applications, such as stab-proof armor, impact mitigation, and smart speed bumps~\cite{Gurgen_2017, Lee_2003}.

In this brief perspective, I will focus on a particular particle size limit (radius $R \gtrsim 1 \mu$m), where Brownian forces can be neglected. The viscosity (ratio of shear stress and rate) in this particular limit should theoretically be rate (or stress) independent. 
However, under shear, a variety of striking features such as yielding, shear-thinning, shear thickening, or even jamming are observed~\cite{Morris_2020, Ness_2022}.
These behaviors ultimately stem from detailed interparticle interactions that are further influenced by solid-fluid interfacial chemistry, roughness, properties of surrounding liquid, etc. 
Given the plethora of factors that can affect these interactions, establishing a predictive link between microscopic particle-level details and the material response is still a daunting task.

\begin{figure}[h]%
\centering
\includegraphics[width=0.9\textwidth]{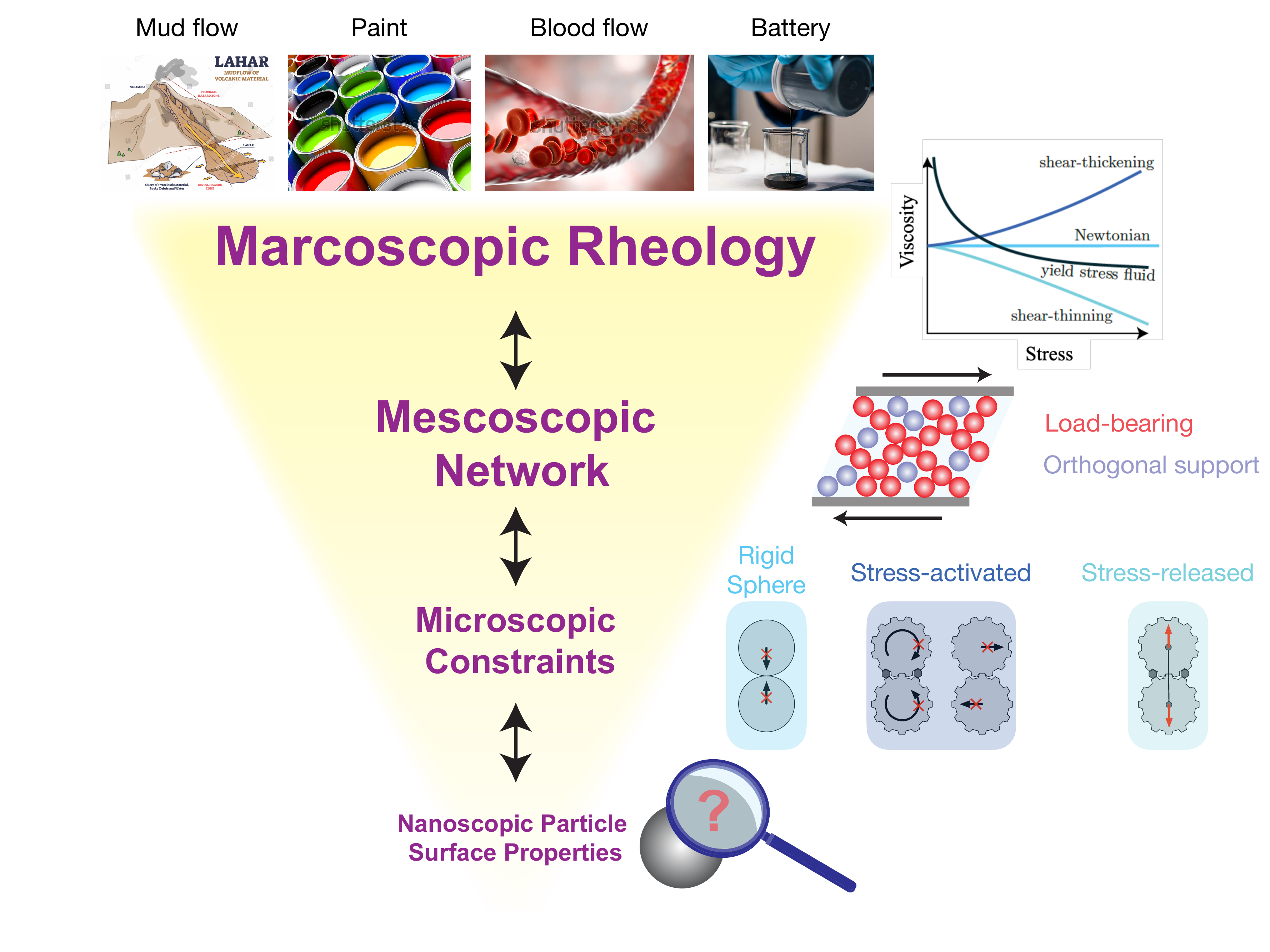}
\caption{\textbf{Hierarchy in the rheology of dense suspensions}: Rheology of dense suspensions can be understood across a hierarchy of length scales and unraveling the interrelationship between macroscopic rheology, mesoscale structural correlations, microscopic constraints hindering relative motion between particles, and nanoscopic particle surface properties.
\textit{Top panel} depicting some relevant practical examples of dense suspensions.}\label{fig1}
\end{figure}

Figure~\ref{fig1} presents the essential challenge associated with creating a link between particle properties and bulk response. 
There exists a hidden hierarchy associated with time-and-length scales that muddle the direct correlation between particle physics and material response.
The particle-level properties such as shape, size, roughness, chemistry, etc. affect the interparticle constraints, viz., sliding, rolling, and twisting.
These constraints on interparticle relative motion further affect the force network that is formed under the action of external deformation.
The formation or breakage of this network ultimately dictates the bulk response or \textit{rheology} of the material.

In what follows, I will introduce each relevant physics for a particular hierarchy and briefly discuss the recent advances. I will eventually close this perspective with the challenges and future issues.

\textbf{Relevant definitions}\label{def}\\
\textit{Constituents}: The particles considered here are mostly rigid, arbitrarily shaped,  having a crystalline or amorphous structure, while in some cases, soft particles are also considered. 
The background solvent mediates the hydrodynamic interactions upon the relative motion of particles. It is mostly a simple liquid (like water), but it can also be polymeric or even a multiphase liquid (cement phase in concrete). Here, we only consider Newtonian solvent with viscosity $\eta_0$.

\textit{Continuum quantities}: Stress $\Sigma$ is a tensorial quantity defined as force per unit area. The full stress tensor can be expressed in terms of shear stress $\sigma = \Sigma_{12}$, particle pressure $\Pi = (\Sigma_{11}+\Sigma_{22}+\Sigma_{33})/3$, first normal stress difference $N_1 = \Sigma_{11} - \Sigma_{11}$, and second normal stress difference $N_2 = \Sigma_{22} - \Sigma_{33}$.
The strain rate tensor $E$ is the symmetric part of the velocity gradient tensor $E = (\nabla u + \nabla u^T)/2$. The strain rate is denoted by $\dot{\gamma}$.
Eventually, the viscosity is defined as $\eta = \sigma/\dot{\gamma}$ and can be interpreted as the resistance offered by the material in response to the external deformation.

\textit{General rheological features}: The term ``Rheology'' was invented by Prof. Bingham and it means \textit{the study of deformation and flow of matter}~\cite{Barnes_1989}.
When the stress increases linearly with the applied deformation rate, $\sigma \propto \dot{\gamma}$ means the viscosity of the system $\eta$ is independent of the deformation rate $\dot{\gamma}$. This viscosity has a unique value at every concentration and increases with increasing $\phi = NV_p/V$ (for $N$ particles of volume $V_p$ in system of volume $V$), often diverging at the jamming transition $\phi_J$~\cite{Morris_2020}.
The precise value of $\phi_J$ depends on particle properties such as their size distribution and their surface interactions.
Often the suspensions show rather complex nonlinear rheological behaviors such as a shear-rate dependence of viscosity, the existence of minimum stress for the suspension to flow (yielding), or time dependence of viscosity (thixotropy or viscoelasticity).
When the viscosity of material $\eta$ decreases with an increase in deformation rate $\dot{\gamma}$ is called shear-thinning.
In many suspensions, the viscosity increases with shear rate (or shear stress). This is termed shear thickening and is the main focus of this perspective.
When viscosity $\eta$ increases continuously with the shear rate $\dot{\gamma}$, it is termed continuous shear thickening (CST).
In some cases, the increase in viscosity can increase abruptly by orders of magnitude and is called discontinuous shear thickening (DST).

\section{Bulk response: connecting rheology with constraints}\label{rheology}
Here, we focus on the simplest case of non-Brownian, neutrally buoyant, rigid particles in a Newtonian fluid under simple shear, i.e., the case where the shear rate is spatially uniform.
The basic ingredients involved in this problem are particle size $R$, solvent viscosity $\eta_0$, and density (equal for solid particles and solvent $\rho_P = \rho_s$).
In the case of rigid particles, there is no force/energy scale.
The macroscopic control variables in the problem are the system size $L$ (here we consider $L \gg R$), solid fraction of particles $\phi$ (relative volume occupied by particles), and strain ($\gamma = \dot{\gamma}t$, $\gamma \to \infty$ considered here). The simple dimensional analysis will imply 4 dimensionless parameters will control the physics here. These are the Stokes number $St \equiv \rho_P \dot{\gamma}R^2/\eta_0$, the Reynolds number $Re \equiv \rho_s \dot{\gamma}L^2/\eta_0$, relative viscosity of the suspension $\eta_r = \eta/\eta_0 = \sigma/\eta_f\dot{\gamma}$, and strain $\gamma$.
The particle size and other details imply that we are in the limit of $St=0$ and $Re=0$. This means that the rheology in steady-state depends only on volume fraction $\phi$ (note that we assumed $\gamma \to \infty$), implying stress is linear in the deformation rate, i.e., Newtonian rheology.
The assumption of the linear relation between $\sigma (\dot{\gamma})$ can in principle be extended to particle pressure $\Pi$ and normal stress differences (mostly found to be negative).
These terms only depend on $\phi$, and in the case of dense suspensions the viscosity (defined for each stress component) eventually diverges at the so-called jamming point as $(\phi_J-\phi)^{-\beta}$ ($\beta$ is mostly found to be 2, but this detail is not important for the discussion).
Jamming behavior is a well-studied concept in dry granular materials~\cite{LiuNagel_AnnRev, Hecke_2009,OHern_2003}, but a rather new one in the case of dense suspensions.
Traditionally the rheology in dense suspension has been approached from a constant volume perspective~\cite{Mewis_2011}. Recent developments have considered pressure-imposed perspectives as well and have demonstrated the rheology of the two cases to be equivalent~\cite{Boyer_2011,Etcheverry_2023,Athani_2022,Clavaud_2020,Dong_2017}.

The dimensional analysis suggests the rheology to be Newtonian, while experimental evidence suggests the rheology to be a combination of Newtonian, shear thinning, and shear thickening/jamming~\cite{Morris_2020, Ness_2022, Brown_2014}.
Newtonian rheology requires that the force exerted on a particle due to external shear must be balanced by another force. 
Experimentally, while preparing the initial sample the particles need to be stabilized against clustering/aggregating using polymer/addition of salt leading to a finite-range repulsive force.
This repulsive force leads to a characteristic repulsive stress $\sigma^* = F_R/R^2$ and competes with the viscous stress $\propto \eta_0\dot{\gamma}$.

The recent surge in research activity has demonstrated that the thickening (or thinning) behavior can be understood in terms of \textit{stress-activated} (or released) constraints on relative motion~\cite{Guy_2018, Singh_2020, Singh_2022}.
In particular, the strong CST or DST has been related to the stress-activated transition from an unconstrained lubricated to a constrained state where the relative motion of particles in the tangential pairwise direction is significantly constrained (or hindered) beyond the onset stress $\sigma^*$~\cite{Seto_2013, Mari_2014, Wyart_2014, Morris_2020, Singh_2020}.
This constraint can originate from direct frictional contact between particles~\cite{Seto_2013, Mari_2014, mari_discontinuous_2015, Ness_2016, Clavaud_2017,Hsu_2018} or enhanced lubrication force at the particle roughness (or asperity) level~\cite{Jamali_2019}.
At stress levels $\sigma \ll \sigma^*$, the lubrication layer between particles is maintained, limiting the role of surface details (like asperity, roughness, etc.) leading to frictionless rheology.
With the increase in applied stress $\sigma$, the repulsive barrier is overcome and lubrication/repulsion gives way to a frictional contact state.
Maxwellian stability argument suggests that the presence of constraints (or friction) leads to a reduction in the number of contacts $Z_c$ needed for mechanical equilibrium (or the iso-static condition); implying $\phi_J^\mu < \phi_J^0$~\cite{Maxwell_1864, Santos_2020, LiuNagel_AnnRev, Hecke_2009}.
With increasing stress, the system makes a transition from a frictionless limit ($\phi_J^0 \equiv \phi_J (\sigma/\sigma^* \to 0)$) to a frictional limit ($\phi_J^\mu \equiv \phi_J (\sigma/\sigma^* \to \infty)$) with $\phi_J^\mu < \phi_J^0$.
This implies that for a system under constant $\phi$, increasing stress $\sigma$ brings the system closer to the jamming behavior.

\begin{figure}[h]%
\centering
\includegraphics[width=0.9\textwidth]{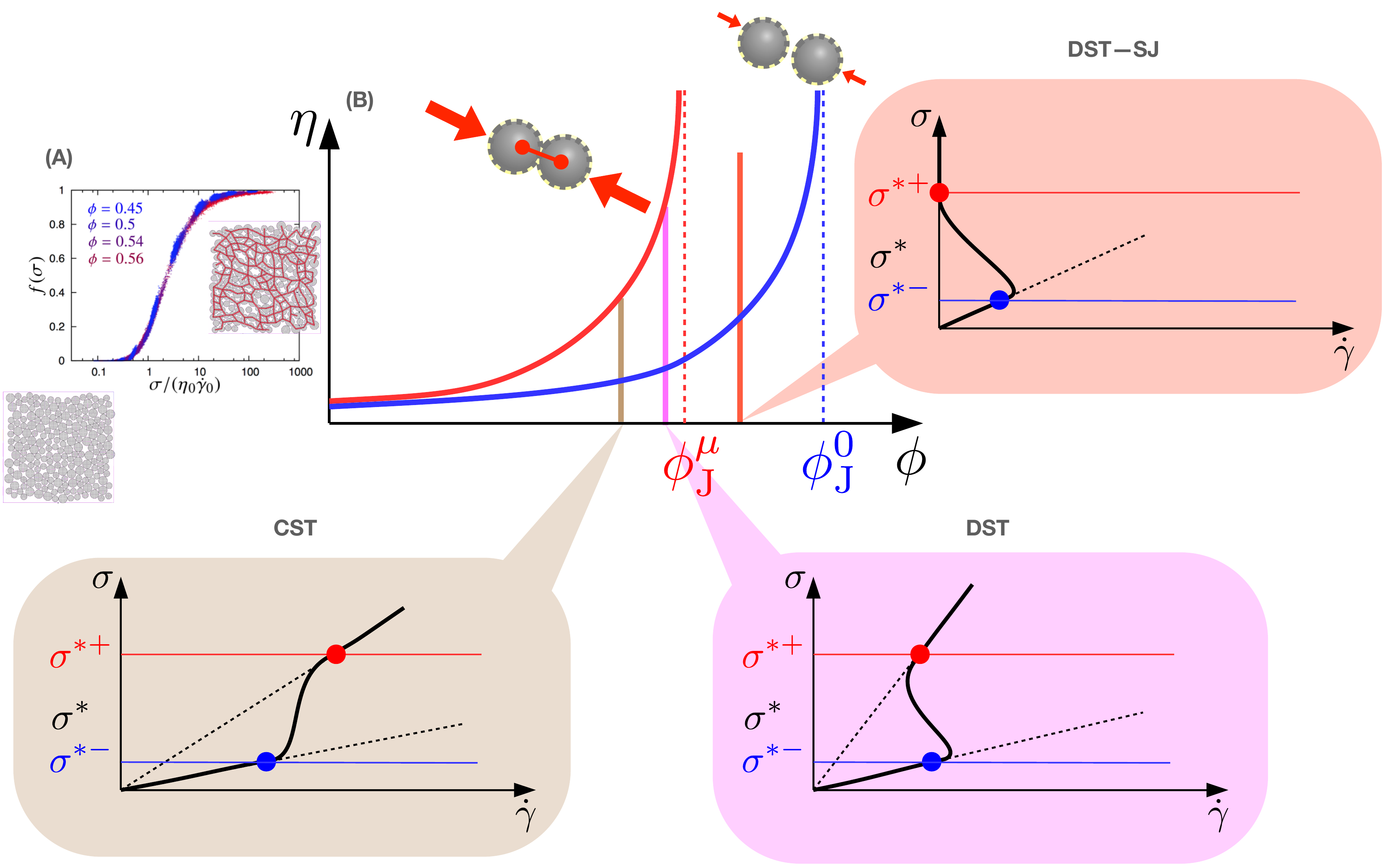}
\caption{\textbf{Shear thickening behavior in dense suspension}. (A) The functional form of the fraction of frictional contacts increasing smoothly from zero to one with increasing stress (from Mari et al.~\cite{Mari_2014}). (B) Two branches of Newtonian viscosity: lower (lubricated, unconstrained frictionless state, with friction coefficient $\mu=0$ leading to $\phi_J^0\approx 0.65$) and upper (constrained, frictional state with friction $\mu>0$ leading to $\phi_J^\mu< \phi_J^0$). Various colored arrows indicate the transitions from unconstrained to constrained state leading to: CST (monotonic $\sigma(\dot{\gamma})$ relation) at low $\phi$, DST (non-monotonic $\sigma(\dot{\gamma})$ relation) at volume fractions approaching but lower than $\phi_J^\mu$), and DST-SJ at volume fractions $\phi>\phi_J^\mu$ (the curve bending back to zero shear rates at large stress).}\label{fig2}
\end{figure}

The mechanism for ``frictional shear thickening'' behavior is illustrated in Fig.~\ref{fig2}.
The essential ingredients are two Newtonian states with distinct jamming points, $\phi_J^0$ and $\phi_J^\mu<\phi_J^0$, and stress-activated particle level constraints. 
It is the activation of constraints on relative motion (e.g. sliding and rolling frictions) that essentially leads to the reduction in the jamming point.
The exact number of contacts $Z$ (or constraints) is rather unimportant, it is rather the degree of constraints (e.g., fraction of frictional contacts $f(\sigma)$ that varies between 0 to 1~\ref{fig2} (A))~\cite{Wyart_2014, Mari_2014}.
This function has usually been reported to be of a sigmoidal shape around the stress needed to form frictional contacts $\sigma^*$, leading to $f(\sigma)\ll \sigma^* \to 0$ and $f(\sigma)\gg \sigma^* \to 1$. This essentially means suspension is in a frictionless state (with viscosity diverging at $\phi_J^0$) when $f(\sigma) \to 0$ and makes a transition to a frictional state 
(with viscosity diverging at $\phi_J^\mu$) when $f(\sigma) \to 1$
The model that rheology is Newtonian for $\sigma \ll \sigma_{-}^*$ and $\sigma \gg \sigma_{+}^*$ and the crossover from frictionless Newtonian state to frictional one results in shear thickening.
The extent of shear thickening depends on the proximity to the frictional jamming point $\phi_J^\mu$ or, in other words, the difference between the viscosity of the two Newtonian states.
When this difference is small for $\phi \ll \phi_J^\mu$, the curve joining two Newtonian states is monotonic, leading to continuous thickening (CST). 
At solid concentrations, $\phi_C \ge \phi < \phi_J^\mu$, the difference between the two states is large but finite, and the flow curve becomes nonmonotonic or S-shaped, thus thickening becomes discontinuous (DST).
At solid concentrations $\phi \ge \phi_J^\mu$, the material makes the more extreme transition from liquid to solid as $\sigma$ increases: known as shear jamming (SJ).
Wyart \& Cates and others have formulated the above philosophy into constitutive models for rate-dependent rheology that have been successfully compared with experiments and numerical simulations~\cite{Wyart_2014, Mari_2014, Guy_2015, Singh_2018, Singh_2020, More_2020, Pradeep_2021}. 

The close surface interactions between particles seem to be important in developing constitutive models for dense suspensions, hence short-range pairwise attraction/repulsion are critical.
%
%
In the case of non-Brownian particles, attractive force originating from van der Waals~\cite{Maranzano_2001}, depletion forces due to dissolved non-interacting polymer~\cite{Gopalakrishnan_2004}, or the presence of an external field~\cite{Brown_2010} can also be present.
Like the case of well-studied colloidal gels~\cite{Larson_1999, Zaccarelli_2007,Mewis_2011}, attraction can introduce competition between the formation of aggregates (sticking of particles) and their eventual breakage (due to shear).
The presence of attractive interactions between particles often leads to the emergence of yield stress (the exact value of which is protocol dependent~\cite{Richards_2020}).
Recent numerical works have included the physics of yielding and subsequent strong shear thinning behavior into the frictional shear thickening framework and postulated more generalized rheological models~\cite{Singh_2018, Guy_2018}.

\section{Connecting mesoscale network with rheology}\label{network}
The deformation-induced solidification or shear jamming (SJ) is common in both granular materials (quasistatic condition) and dense suspensions (Stokes-flow).
Experiments from the late Prof. Behringer's group have demonstrated the appearance of a load-bearing network for mm-sized photoelastic particles under shear.
These systems would jam under shear while staying in a liquid-like state under isotropic conditions.
Thus shearing leads to a self-organized jammed state to support the external stress.
Shear jamming is thus closely related to the collective organization in the space of forces (this space is dual to the real particle space).
The network formed under shear becomes strongly correlated because of 
the constraints of force and torque balance on every grain.
Cates \textit{et al.}~\cite{Cates_1998a} suggested that such a jammed (SJ) state is formed principally by the primary force chain along the compression axis with support from secondary force chains along the orthogonal tensile direction. They also postulated that such a jammed state is fragile, in the sense that this state is maintained only by the imposed load, and would fail if the load is removed, or applied in the reverse direction.
Radjai~ \textit{et al.}~\cite{Radjai_1998} showed the existence of two complementary networks in dry granular systems under shear: load-bearing percolating ``strong force chain'' network oriented along the major (compressive) axis along with a network of ``weak force chains'' in the orthogonal direction. 
Chakraborty and coworkers have developed theoretical tools to identify the force-space organization that has been instrumental in understanding shear jamming~\cite{Sarkar_2013, Sarkar_2015, Sarkar_2016, thomas2018microscopic}.

DST and SJ have been associated with the load-bearing frictional network formed under shear in dense suspensions.
This load-bearing system spanning network resists large deformation leading to sudden orders of magnitude increase in viscosity.
Strikingly, the classical theoretical tools like pair-correlation functions of particle-level dynamics $g(r), S(k)$ do not display the distinct difference between thickened and unthickened states, while their rheology is distinctly different with orders of magnitude difference in the viscosity (along with often change in sign of first normal stress difference $N_1$~\cite{Royer_2016})~\cite{Mari_2014}.
This prompted the use of network science-related tools to understand the microscopic (or rather mesoscale) reasoning for strong shear thickening or jamming.
However, these studies are limited compared to the dry granular materials, partly because the concept of contact (enduring) between particles is rather new to dense suspensions and is still being debated~\cite{Jamali_2020, Lee_2020}.
Mari \textit{et al.}~\cite{Mari_2014} showed that percolating frictional force chains with orthogonal support form at the volume fractions close to DST, which is qualitatively similar to ideas put forth by Cates \textit{et al.} in dry granular materials~\cite{Cates_1998a}.
Further, Boromand \textit{et al.}~\cite{Boromand_2018} analyzed these networks in terms of giant clusters and showed correspondence between DST and giant cluster growth.
Gameiro \textit{et al.}~\cite{Gameiro_2020} used topological tools called Persistence homology to connect the loop-like structure growth (the minimally rigid structure that can resist simple shear deformation) to strong shear thickening behavior.
Edens \textit{et al.}~\cite{Edens_2021} further investigated correlations between the geometric organization of particles, underlying forces, and rheological properties in simulations where frictional shear thickening behavior was coupled with a strong attraction that leads to yielding behavior~\cite{Singh_2019}. The authors demonstrated that the changes in suspension rheology (from thinning to thickening) do not simply originate from local/global-scale particle rearrangements. These changes rather sensitively depend on the detailed balance of forces, resulting force network, and are coupled with the external deformation strength (shear stress in this case).
Nabizadeh \textit{et al.}~\cite{Nabizadeh_2022} have analyzed the network using the community detection algorithm and suggested that DST is correlated with the enhanced constraints on the relative motion between clusters.
Rather recent studies have demonstrated the system-spanning enduring rigid clusters to be responsible for DST~\cite{Naald_2023, Goyal_2022}.

\section{Connecting constraints with particle-level details: tuning \& manipulating the rheology}\label{constraints}
\begin{figure}[h]%
\centering
\includegraphics[width=0.9\textwidth]{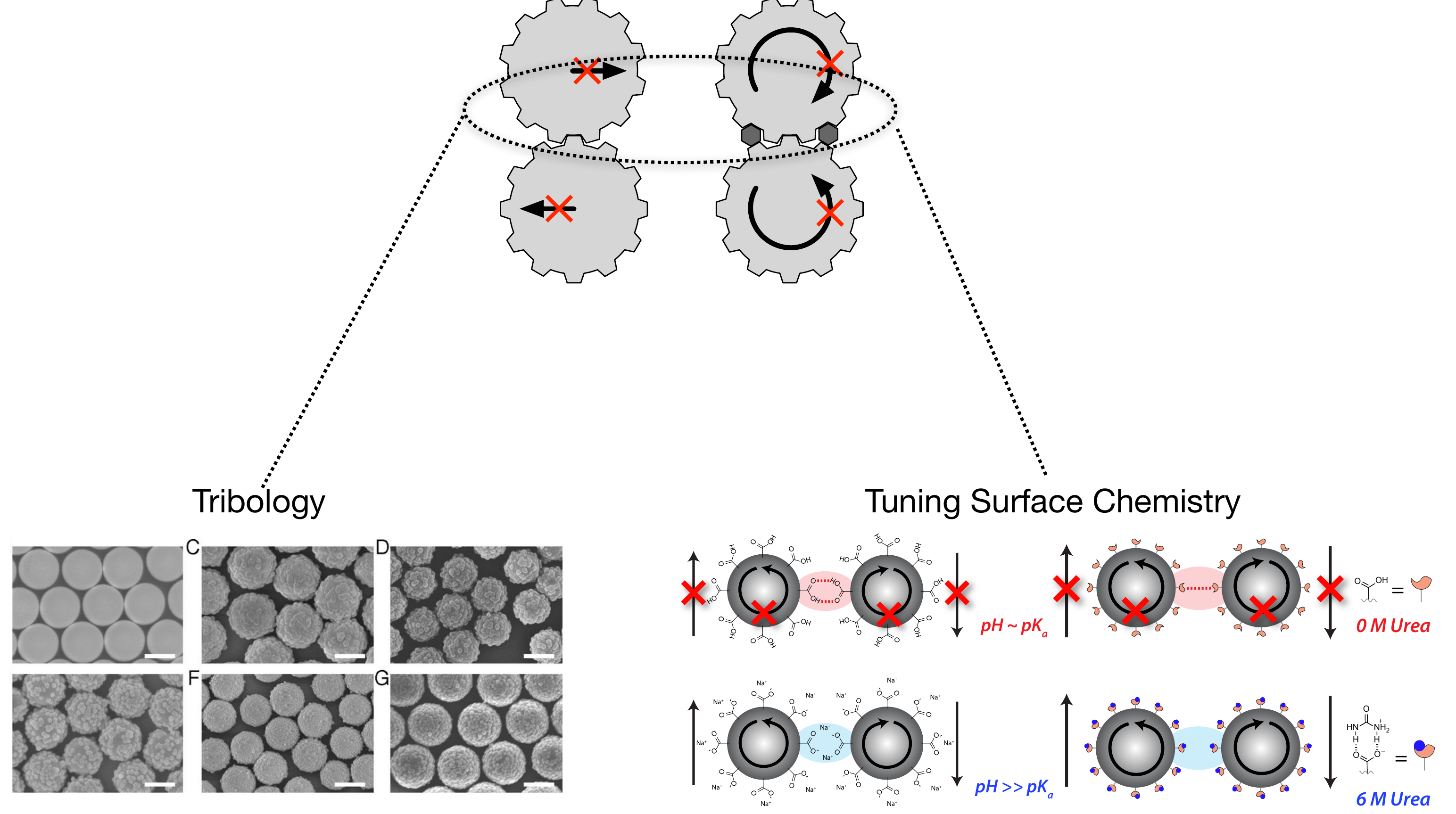}
\caption{\textbf{Stress-activated constraints originating from particle level details}. Both particle roughness and interfacial chemistry can modify both sliding and rolling friction between particles. (Left): Increasing the asperity size can lead to their interlocking and thus enhancing friction (especially rolling friction). (Right): Addition of urea to carboxylic acid or decreasing the pH of solution containing carboxylic acid coated particles can both lead to disruption of hydrogen bonding further reduce the constraints.
}\label{fig3}
\end{figure}

The description in Sec.~\ref{rheology} and thereby mean-field models are sensitive to the exact value of jamming volume friction $\phi_J^\mu$.
The measurement of forces between particles of size 100 nm to 100 $\mu$m is rather difficult experimentally, though some progress has been made recently~\cite{Comtet_2017}.
A rather different approach is to focus on the general types of constraints that can hinder the relative motion between particles~\cite{Guy_2018, Singh_2020, Singh_2022, Singh_2022a}.
The idea is given the frictional contact between two particles can hinder sliding only, or both sliding as well as rolling motion with sliding and rolling friction parameters ($\mu_s$ and $\mu_r$, respectively).
The promise of this approach is that the chemical or physical origin on a particular type of interparticle interaction might matter far less as compared to the constraint that it offers.
Singh \textit{et al.}~\cite{Singh_2022, Singh_2022a} have recently shown that this constraint-based approach can quantitatively reproduce shear thickening behavior across diverse kinds of dense suspensions that include a wide range of particle sizes and surface features.
This work offers a quantitive understanding of shear thickening behavior in dense suspensions -- a deep understanding that paves the way to tune and manipulate the rheology for practical purposes.

A number of previous experimental studies have considered possible surface effects (including modifying surface or solvent-particle interactions) thus affecting constraints between relative particle motion~\ref{fig3}.
Here, in this perspective, I have only included a few of these studies (readers are encouraged to check other rather more detailed reviews~\cite{Morris_2020, Ness_2022}.
A series of studies~\cite{Castle_1996, Lootens_2003, Lootens_2004} have shown that roughness at particle level can drastically influence the rheology: leading to enhanced viscosity, reduction in onset stress for shear thickening, and change in sign of the first normal stress difference $N_1$.
More recently Hsu \textit{et al.}~\cite{Hsu_2018} and Hsiao \textit{et al.}~\cite{Hsiao_2017} have shown a systematic reduction in $\phi_J^\mu$ with increasing roughness along with an increase in stress range over which shear thickening occurs.

Researchers have also used particle and particle-fluid interfacial chemistry to tune shear thickening behavior.
James \textit{et al.}~\cite{james2018interparticle} have shown that the addition of urea can decrease shear thickening for particles coated with carboxylic acid group (-CO$_2$H) dispersed in an aqueous solution.
In a subsequent study~\cite{James_2019}, disruption of hydrogen bonding by ``capping'' of the (-CO$_2$H) was demonstrated to be responsible for this reduction.
Tuning pH of the solvent~\cite{Laun_1984}, the molecular weight of the polymeric solvent~\cite{Xu_2020, Naald_2021}, and more recently using micron-sized particles with accessible glass transition temperatures~\cite{Chen_2022} are ways to tune the constraints and thus manipulate the rheology.

\section{Closing remarks}\label{remarks}
This perspective aims to outline the current understanding of the rheology of suspension dynamics and present a vision of the hierarchy of length and time scales that are involved. 
Regarding practical relevance, $\phi > 0.5$ can be considered dense. However, the exact number depends on the distance from frictional jamming point $\phi_J^\mu$ that depends on surface features or constraints ($\mu_s, \mu_r$)~\cite{Singh_2022}.
The rapid recent development of the physics of dense suspensions has put experiments, simulations, and theory on the same footing.
However, this understanding is limited to the ``ideal" model suspensions (nearly monodisperse rigid particles in Newtonian solvent).
Extending the simple microscopic arguments and the models developed here to continuum descriptions of ``real'' world suspensions will be a nontrivial task but is needed to tackle these pressing challenges.
Particles in the real world are rough, non-spherical, sticky, and polydisperse, and the background solvent can be non-Newtonian.\\
%

%
%

A few issues/questions that need close attention are:

\begin{itemize}
    \item Can the newly proposed constitutive models for ideal nearly monodisperse spheres be extended to real-world particles-fluid mixtures, i.e., particles with polydispersity in shape and size dispersed in a non-Newtonian solvent?
    \item The current simulation (and theoretical) approaches have considered stress-activated friction, while the influence of strong attraction and adhesion has not been considered in detail. There has been little examination of how the external fields (electrical, magnetic, acoustic) leading to forces e.g., particle polarization in an electric field would affect the fine structure. 
    \item The pressure-controlled boundary condition is most relevant to natural and industrial connections. However, very little attention has been paid to the constitutive modeling of a pressure-controlled setup.
    \item Current algorithms have been highly successful in predicting the dense suspension rheology but are limited to a few thousand particles. Thus, designing clever algorithms dealing with both better contact detection between particles and memory allocation without losing any essential physics is the need of the hour.
    \item Large-scale mesoscale correlations and force network organization lead to large viscosity (close to jamming). Investigation relating microscopic constraints to network topology/geometry can be helpful in building up a statistical mechanics framework for out-of-equilibrium systems.
\end{itemize}

This multi-faceted problem requires experts from soft condensed matter physics, engineering, technology, chemistry, computer science, and network science.
%
Not only do dense suspensions have important practical applications, but there's plenty of room for all scientists to play across this hierarchy of length scales.

\bmhead{Acknowledgments}

I acknowledge Case Western Reserve University for the start-up funding for the project.  

I am grateful to my mentors Heinrich M. Jaeger and Juan J. de Pablo (University of Chicago); Jeffrey F. Morris and Morton M. Denn (Levich Institute, City College of New York); Stefan Luding and Vanessa Magnanimo (University of Twente, The Netherlands) for their guidance that has been critical in shaping my thoughts. 

The work presented here has been performed in several collaborations with Romain Mari, Ryohei Seto, Christopher Ness, Sidhant Pednekar, Jaehun Chun, Grayson L. Jackson, and Michael van der Naald.

I would also like to thank Stuart Rowan, Bulbul Chakraborty, Emanuela del Gado, Safa Jamali, Lilian Hsiao, Lou Kondic, Aurora Clark, Jacinta Konrad, Sarah Hormozi, Douglas Jerolmack, and Karen Daniels for many insightful discussions. I appreciate the collaborations with Omer Sedes, Jetin E Thomas, Kabir Ramola, Qin Xu, Marcio Gameiro, and Konstantin Mischaikow. I have thoroughly enjoyed working with my colleagues and friends: Elise Chen, Endao Han, Nicole M James, Neil Dolinski, Melody X Lim, and Bryan VanSaders for many fruitful discussions over the years.

\section*{Declarations}

\begin{itemize}
\item Funding: \textit{Not applicable} 
\item Conflict of interest/Competing interests (check journal-specific guidelines for which heading to use): \textit{None}
\item Ethics approval 
\item Consent to participate 
\item Consent for publication \textit{Yes}
\item Availability of data and materials
\item Code availability \textit{Yes}
\item Authors' contributions \textit{Not applicable}
\end{itemize}


\bibliography{dst}


\begin{thebibliography}{72}
\ifx \bisbn   \undefined \def \bisbn  #1{ISBN #1}\fi
\ifx \binits  \undefined \def \binits#1{#1}\fi
\ifx \bauthor  \undefined \def \bauthor#1{#1}\fi
\ifx \batitle  \undefined \def \batitle#1{#1}\fi
\ifx \bjtitle  \undefined \def \bjtitle#1{#1}\fi
\ifx \bvolume  \undefined \def \bvolume#1{\textbf{#1}}\fi
\ifx \byear  \undefined \def \byear#1{#1}\fi
\ifx \bissue  \undefined \def \bissue#1{#1}\fi
\ifx \bfpage  \undefined \def \bfpage#1{#1}\fi
\ifx \blpage  \undefined \def \blpage #1{#1}\fi
\ifx \burl  \undefined \def \burl#1{\textsf{#1}}\fi
\ifx \doiurl  \undefined \def \doiurl#1{\url{https://doi.org/#1}}\fi
\ifx \betal  \undefined \def \betal{\textit{et al.}}\fi
\ifx \binstitute  \undefined \def \binstitute#1{#1}\fi
\ifx \binstitutionaled  \undefined \def \binstitutionaled#1{#1}\fi
\ifx \bctitle  \undefined \def \bctitle#1{#1}\fi
\ifx \beditor  \undefined \def \beditor#1{#1}\fi
\ifx \bpublisher  \undefined \def \bpublisher#1{#1}\fi
\ifx \bbtitle  \undefined \def \bbtitle#1{#1}\fi
\ifx \bedition  \undefined \def \bedition#1{#1}\fi
\ifx \bseriesno  \undefined \def \bseriesno#1{#1}\fi
\ifx \blocation  \undefined \def \blocation#1{#1}\fi
\ifx \bsertitle  \undefined \def \bsertitle#1{#1}\fi
\ifx \bsnm \undefined \def \bsnm#1{#1}\fi
\ifx \bsuffix \undefined \def \bsuffix#1{#1}\fi
\ifx \bparticle \undefined \def \bparticle#1{#1}\fi
\ifx \barticle \undefined \def \barticle#1{#1}\fi
\bibcommenthead
\ifx \bconfdate \undefined \def \bconfdate #1{#1}\fi
\ifx \botherref \undefined \def \botherref #1{#1}\fi
\ifx \url \undefined \def \url#1{\textsf{#1}}\fi
\ifx \bchapter \undefined \def \bchapter#1{#1}\fi
\ifx \bbook \undefined \def \bbook#1{#1}\fi
\ifx \bcomment \undefined \def \bcomment#1{#1}\fi
\ifx \oauthor \undefined \def \oauthor#1{#1}\fi
\ifx \citeauthoryear \undefined \def \citeauthoryear#1{#1}\fi
\ifx \endbibitem  \undefined \def \endbibitem {}\fi
\ifx \bconflocation  \undefined \def \bconflocation#1{#1}\fi
\ifx \arxivurl  \undefined \def \arxivurl#1{\textsf{#1}}\fi
\csname PreBibitemsHook\endcsname

\bibitem[\protect\citeauthoryear{Denn et~al.}{2018}]{Denn_2018}
\begin{barticle}
\bauthor{\bsnm{Denn}, \binits{M.M.}},
\bauthor{\bsnm{Morris}, \binits{J.F.}},
\bauthor{\bsnm{Bonn}, \binits{D.}}:
\batitle{Shear thickening in concentrated suspensions of smooth spheres in
  newtonian suspending fluids}.
\bjtitle{Soft Matter}
\bvolume{14}(\bissue{2}),
\bfpage{170}--\blpage{184}
(\byear{2018})
\end{barticle}
\endbibitem

\bibitem[\protect\citeauthoryear{Denn and Morris}{2014}]{Denn_2014}
\begin{botherref}
\oauthor{\bsnm{Denn}, \binits{M.M.}},
\oauthor{\bsnm{Morris}, \binits{J.F.}}:
Rheology of non-{B}rownian suspensions.
Annu. Rev. Chem. Biomol. Eng.
\textbf{5}(1)
(2014)
\end{botherref}
\endbibitem

\bibitem[\protect\citeauthoryear{Guazzelli and
  Pouliquen}{2018}]{guazzelli_2018}
\begin{botherref}
\oauthor{\bsnm{Guazzelli}, \binits{{\'E}.}},
\oauthor{\bsnm{Pouliquen}, \binits{O.}}:
Rheology of dense granular suspensions.
J. Fluid Mech.
\textbf{852}
(2018)
\end{botherref}
\endbibitem

\bibitem[\protect\citeauthoryear{Jerolmack and Daniels}{2019}]{Jerolmack_2019}
\begin{barticle}
\bauthor{\bsnm{Jerolmack}, \binits{D.J.}},
\bauthor{\bsnm{Daniels}, \binits{K.E.}}:
\batitle{Viewing earth’s surface as a soft-matter landscape}.
\bjtitle{Nature Reviews Physics}
\bvolume{1}(\bissue{12}),
\bfpage{716}--\blpage{730}
(\byear{2019})
\end{barticle}
\endbibitem

\bibitem[\protect\citeauthoryear{Morris}{2020}]{Morris_2020}
\begin{barticle}
\bauthor{\bsnm{Morris}, \binits{J.F.}}:
\batitle{Shear thickening of concentrated suspensions: Recent developments and
  relation to other phenomena}.
\bjtitle{Annual Review of Fluid Mechanics}
\bvolume{52},
\bfpage{121}--\blpage{144}
(\byear{2020})
\end{barticle}
\endbibitem

\bibitem[\protect\citeauthoryear{Brady and Bossis}{1985}]{Brady_1985}
\begin{barticle}
\bauthor{\bsnm{Brady}, \binits{J.F.}},
\bauthor{\bsnm{Bossis}, \binits{G.}}:
\batitle{The rheology of concentrated suspensions of spheres in simple shear
  flow by numerical simulation}.
\bjtitle{J. Fluid Mech.}
\bvolume{155},
\bfpage{105}--\blpage{129}
(\byear{1985})
\end{barticle}
\endbibitem

\bibitem[\protect\citeauthoryear{Wagner and Brady}{2009}]{Wagner_2009}
\begin{barticle}
\bauthor{\bsnm{Wagner}, \binits{N.J.}},
\bauthor{\bsnm{Brady}, \binits{J.F.}}:
\batitle{Shear thickening in colloidal dispersions}.
\bjtitle{Phys. Today}
\bvolume{62},
\bfpage{27}--\blpage{32}
(\byear{2009})
\end{barticle}
\endbibitem

\bibitem[\protect\citeauthoryear{Barnes et~al.}{1989}]{Barnes_1989}
\begin{bbook}
\bauthor{\bsnm{Barnes}, \binits{H.A.}},
\bauthor{\bsnm{Hutton}, \binits{J.F.}},
\bauthor{\bsnm{Walters}, \binits{K.}}:
\bbtitle{An Introduction to Rheology}
vol. \bseriesno{3}.
\bpublisher{Elsevier}, \blocation{???}
(\byear{1989})
\end{bbook}
\endbibitem

\bibitem[\protect\citeauthoryear{G{\"u}rgen et~al.}{2017}]{Gurgen_2017}
\begin{barticle}
\bauthor{\bsnm{G{\"u}rgen}, \binits{S.}},
\bauthor{\bsnm{Ku{\c{s}}han}, \binits{M.C.}},
\bauthor{\bsnm{Li}, \binits{W.}}:
\batitle{Shear thickening fluids in protective applications: A review}.
\bjtitle{Progress in Polymer Science}
\bvolume{75},
\bfpage{48}--\blpage{72}
(\byear{2017})
\end{barticle}
\endbibitem

\bibitem[\protect\citeauthoryear{Lee et~al.}{2003}]{Lee_2003}
\begin{barticle}
\bauthor{\bsnm{Lee}, \binits{Y.S.}},
\bauthor{\bsnm{Wetzel}, \binits{E.D.}},
\bauthor{\bsnm{Wagner}, \binits{N.J.}}:
\batitle{The ballistic impact characteristics of kevlar{\textregistered} woven
  fabrics impregnated with a colloidal shear thickening fluid}.
\bjtitle{Journal of materials science}
\bvolume{38},
\bfpage{2825}--\blpage{2833}
(\byear{2003})
\end{barticle}
\endbibitem

\bibitem[\protect\citeauthoryear{Ness et~al.}{2022}]{Ness_2022}
\begin{barticle}
\bauthor{\bsnm{Ness}, \binits{C.}},
\bauthor{\bsnm{Seto}, \binits{R.}},
\bauthor{\bsnm{Mari}, \binits{R.}}:
\batitle{The physics of dense suspensions}.
\bjtitle{Annual Review of Condensed Matter Physics}
\bvolume{13},
\bfpage{97}--\blpage{117}
(\byear{2022})
\end{barticle}
\endbibitem

\bibitem[\protect\citeauthoryear{Liu and Nagel}{2010}]{LiuNagel_AnnRev}
\begin{barticle}
\bauthor{\bsnm{Liu}, \binits{A.J.}},
\bauthor{\bsnm{Nagel}, \binits{S.R.}}:
\batitle{The jamming transition and the marginally jammed solid}.
\bjtitle{Annu. Rev. Condens. Matter Phys.}
\bvolume{1}(\bissue{1}),
\bfpage{347}--\blpage{369}
(\byear{2010})
\end{barticle}
\endbibitem

\bibitem[\protect\citeauthoryear{van Hecke}{2009}]{Hecke_2009}
\begin{barticle}
\bauthor{\bsnm{Hecke}, \binits{M.}}:
\batitle{Jamming of soft particles: geometry, mechanics, scaling and
  isostaticity}.
\bjtitle{J. Phys. Condens. Matter}
\bvolume{22}(\bissue{3}),
\bfpage{033101}
(\byear{2009})
\end{barticle}
\endbibitem

\bibitem[\protect\citeauthoryear{O'Hern et~al.}{2003}]{OHern_2003}
\begin{barticle}
\bauthor{\bsnm{O'Hern}, \binits{C.S.}},
\bauthor{\bsnm{Silbert}, \binits{L.E.}},
\bauthor{\bsnm{Liu}, \binits{A.J.}},
\bauthor{\bsnm{Nagel}, \binits{S.R.}}:
\batitle{Jamming at zero temperature and zero applied stress: The epitome of
  disorder}.
\bjtitle{Phys. Rev. E}
\bvolume{68},
\bfpage{011306}
(\byear{2003})
\end{barticle}
\endbibitem

\bibitem[\protect\citeauthoryear{Mewis and Wagner}{2011}]{Mewis_2011}
\begin{bbook}
\bauthor{\bsnm{Mewis}, \binits{J.}},
\bauthor{\bsnm{Wagner}, \binits{N.J.}}:
\bbtitle{Colloidal Suspension Rheology}.
\bpublisher{Cambridge University Press}, \blocation{???}
(\byear{2011})
\end{bbook}
\endbibitem

\bibitem[\protect\citeauthoryear{Boyer et~al.}{2011}]{Boyer_2011}
\begin{barticle}
\bauthor{\bsnm{Boyer}, \binits{F.}},
\bauthor{\bsnm{Guazzelli}, \binits{{\'E}.}},
\bauthor{\bsnm{Pouliquen}, \binits{O.}}:
\batitle{Unifying suspension and granular rheology}.
\bjtitle{Phys. Rev. Lett.}
\bvolume{107},
\bfpage{188301}
(\byear{2011})
\end{barticle}
\endbibitem

\bibitem[\protect\citeauthoryear{Etcheverry et~al.}{2023}]{Etcheverry_2023}
\begin{barticle}
\bauthor{\bsnm{Etcheverry}, \binits{B.}},
\bauthor{\bsnm{Forterre}, \binits{Y.}},
\bauthor{\bsnm{Metzger}, \binits{B.}}:
\batitle{Capillary-stress controlled rheometer reveals the dual rheology of
  shear-thickening suspensions}.
\bjtitle{Physical Review X}
\bvolume{13}(\bissue{1}),
\bfpage{011024}
(\byear{2023})
\end{barticle}
\endbibitem

\bibitem[\protect\citeauthoryear{Athani et~al.}{2022}]{Athani_2022}
\begin{barticle}
\bauthor{\bsnm{Athani}, \binits{S.}},
\bauthor{\bsnm{Metzger}, \binits{B.}},
\bauthor{\bsnm{Forterre}, \binits{Y.}},
\bauthor{\bsnm{Mari}, \binits{R.}}:
\batitle{Transient flows and migration in granular suspensions: key role of
  reynolds-like dilatancy}.
\bjtitle{Journal of Fluid Mechanics}
\bvolume{949},
\bfpage{9}
(\byear{2022})
\end{barticle}
\endbibitem

\bibitem[\protect\citeauthoryear{Clavaud et~al.}{2020}]{Clavaud_2020}
\begin{barticle}
\bauthor{\bsnm{Clavaud}, \binits{C.}},
\bauthor{\bsnm{Metzger}, \binits{B.}},
\bauthor{\bsnm{Forterre}, \binits{Y.}}:
\batitle{The darcytron: A pressure-imposed device to probe the frictional
  transition in shear-thickening suspensions}.
\bjtitle{Journal of Rheology}
\bvolume{64}(\bissue{2}),
\bfpage{395}--\blpage{403}
(\byear{2020})
\end{barticle}
\endbibitem

\bibitem[\protect\citeauthoryear{Dong and Trulsson}{2017}]{Dong_2017}
\begin{barticle}
\bauthor{\bsnm{Dong}, \binits{J.}},
\bauthor{\bsnm{Trulsson}, \binits{M.}}:
\batitle{Analog of discontinuous shear thickening flows under confining
  pressure}.
\bjtitle{Phys. Rev. Fluids}
\bvolume{2}(\bissue{8}),
\bfpage{081301}
(\byear{2017})
\end{barticle}
\endbibitem

\bibitem[\protect\citeauthoryear{Brown and Jaeger}{2014}]{Brown_2014}
\begin{barticle}
\bauthor{\bsnm{Brown}, \binits{E.}},
\bauthor{\bsnm{Jaeger}, \binits{H.M.}}:
\batitle{Shear thickening in concentrated suspensions: phenomenology,
  mechanisms and relations to jamming}.
\bjtitle{Rep. Prog. Phys.}
\bvolume{77}(\bissue{4}),
\bfpage{046602}
(\byear{2014})
\end{barticle}
\endbibitem

\bibitem[\protect\citeauthoryear{Guy et~al.}{2018}]{Guy_2018}
\begin{barticle}
\bauthor{\bsnm{Guy}, \binits{B.M.}},
\bauthor{\bsnm{Richards}, \binits{J.}},
\bauthor{\bsnm{Hodgson}, \binits{D.}},
\bauthor{\bsnm{Blanco}, \binits{E.}},
\bauthor{\bsnm{Poon}, \binits{W.C.K.}}:
\batitle{Constraint-based approach to granular dispersion rheology}.
\bjtitle{Phys. Rev. Lett.}
\bvolume{121}(\bissue{12}),
\bfpage{128001}
(\byear{2018})
\end{barticle}
\endbibitem

\bibitem[\protect\citeauthoryear{Singh et~al.}{2020}]{Singh_2020}
\begin{barticle}
\bauthor{\bsnm{Singh}, \binits{A.}},
\bauthor{\bsnm{Ness}, \binits{C.}},
\bauthor{\bsnm{Seto}, \binits{R.}},
\bauthor{\bsnm{Pablo}, \binits{J.J.}},
\bauthor{\bsnm{Jaeger}, \binits{H.M.}}:
\batitle{Shear thickening and jamming of dense suspensions: The ``roll'' of
  friction}.
\bjtitle{Phys. Rev. Lett.}
\bvolume{124},
\bfpage{248005}
(\byear{2020})
\end{barticle}
\endbibitem

\bibitem[\protect\citeauthoryear{Singh et~al.}{2022}]{Singh_2022}
\begin{barticle}
\bauthor{\bsnm{Singh}, \binits{A.}},
\bauthor{\bsnm{Jackson}, \binits{G.L.}},
\bauthor{\bsnm{Naald}, \binits{M.}},
\bauthor{\bsnm{Pablo}, \binits{J.J.}},
\bauthor{\bsnm{Jaeger}, \binits{H.M.}}:
\batitle{Stress-activated constraints in dense suspension rheology}.
\bjtitle{Physical Review Fluids}
\bvolume{7}(\bissue{5}),
\bfpage{054302}
(\byear{2022})
\end{barticle}
\endbibitem

\bibitem[\protect\citeauthoryear{Seto et~al.}{2013}]{Seto_2013}
\begin{barticle}
\bauthor{\bsnm{Seto}, \binits{R.}},
\bauthor{\bsnm{Botet}, \binits{R.}},
\bauthor{\bsnm{Meireles}, \binits{M.}},
\bauthor{\bsnm{Auernhammer}, \binits{G.K.}},
\bauthor{\bsnm{Cabane}, \binits{B.}}:
\batitle{Compressive consolidation of strongly aggregated particle gels}.
\bjtitle{J. Rheol.}
\bvolume{57}(\bissue{5}),
\bfpage{1347}--\blpage{1366}
(\byear{2013})
\end{barticle}
\endbibitem

\bibitem[\protect\citeauthoryear{Mari et~al.}{2014}]{Mari_2014}
\begin{barticle}
\bauthor{\bsnm{Mari}, \binits{R.}},
\bauthor{\bsnm{Seto}, \binits{R.}},
\bauthor{\bsnm{Morris}, \binits{J.F.}},
\bauthor{\bsnm{Denn}, \binits{M.M.}}:
\batitle{Shear thickening, frictionless and frictional rheologies in
  non-{B}rownian suspensions}.
\bjtitle{J. Rheol.}
\bvolume{58}(\bissue{6}),
\bfpage{1693}--\blpage{1724}
(\byear{2014})
\end{barticle}
\endbibitem

\bibitem[\protect\citeauthoryear{Wyart and Cates}{2014}]{Wyart_2014}
\begin{barticle}
\bauthor{\bsnm{Wyart}, \binits{M.}},
\bauthor{\bsnm{Cates}, \binits{M.E.}}:
\batitle{Discontinuous shear thickening without inertia in dense non-{B}rownian
  suspensions}.
\bjtitle{Phys. Rev. Lett.}
\bvolume{112},
\bfpage{098302}
(\byear{2014})
\end{barticle}
\endbibitem

\bibitem[\protect\citeauthoryear{Mari et~al.}{2015}]{mari_discontinuous_2015}
\begin{barticle}
\bauthor{\bsnm{Mari}, \binits{R.}},
\bauthor{\bsnm{Seto}, \binits{R.}},
\bauthor{\bsnm{Morris}, \binits{J.F.}},
\bauthor{\bsnm{Denn}, \binits{M.M.}}:
\batitle{Discontinuous shear thickening in {Brownian} suspensions by dynamic
  simulation}.
\bjtitle{Proc. Natl. Acad. Sci. U.S.A.}
\bvolume{112}(\bissue{50}),
\bfpage{15326}--\blpage{15330}
(\byear{2015})
\end{barticle}
\endbibitem

\bibitem[\protect\citeauthoryear{Ness and Sun}{2016}]{Ness_2016}
\begin{barticle}
\bauthor{\bsnm{Ness}, \binits{C.}},
\bauthor{\bsnm{Sun}, \binits{J.}}:
\batitle{Shear thickening regimes of dense non-{B}rownian suspensions}.
\bjtitle{Soft Matter}
\bvolume{12}(\bissue{3}),
\bfpage{914}--\blpage{924}
(\byear{2016})
\end{barticle}
\endbibitem

\bibitem[\protect\citeauthoryear{Clavaud et~al.}{2017}]{Clavaud_2017}
\begin{botherref}
\oauthor{\bsnm{Clavaud}, \binits{C.}},
\oauthor{\bsnm{B{\'e}rut}, \binits{A.}},
\oauthor{\bsnm{Metzger}, \binits{B.}},
\oauthor{\bsnm{Forterre}, \binits{Y.}}:
Revealing the frictional transition in shear-thickening suspensions.
Proc. Natl. Acad. Sci. U.S.A.,
5147--5152
(2017)
\end{botherref}
\endbibitem

\bibitem[\protect\citeauthoryear{Hsu et~al.}{2018}]{Hsu_2018}
\begin{botherref}
\oauthor{\bsnm{Hsu}, \binits{C.-P.}},
\oauthor{\bsnm{Ramakrishna}, \binits{S.N.}},
\oauthor{\bsnm{Zanini}, \binits{M.}},
\oauthor{\bsnm{Spencer}, \binits{N.D.}},
\oauthor{\bsnm{Isa}, \binits{L.}}:
Roughness-dependent tribology effects on discontinuous shear thickening.
Proc. Nat. Acad. Sci.
(2018)
\end{botherref}
\endbibitem

\bibitem[\protect\citeauthoryear{Jamali and Brady}{2019}]{Jamali_2019}
\begin{barticle}
\bauthor{\bsnm{Jamali}, \binits{S.}},
\bauthor{\bsnm{Brady}, \binits{J.F.}}:
\batitle{Alternative frictional model for discontinuous shear thickening of
  dense suspensions: Hydrodynamics}.
\bjtitle{Physical review letters}
\bvolume{123}(\bissue{13}),
\bfpage{138002}
(\byear{2019})
\end{barticle}
\endbibitem

\bibitem[\protect\citeauthoryear{Maxwell}{1864}]{Maxwell_1864}
\begin{barticle}
\bauthor{\bsnm{Maxwell}, \binits{J.C.}}:
\batitle{On the calculation of the equilibrium and stiffness of frames}.
\bjtitle{Philos. Mag.}
\bvolume{27}(\bissue{182}),
\bfpage{294}--\blpage{299}
(\byear{1864})
\end{barticle}
\endbibitem

\bibitem[\protect\citeauthoryear{Santos et~al.}{2020}]{Santos_2020}
\begin{barticle}
\bauthor{\bsnm{Santos}, \binits{A.P.}},
\bauthor{\bsnm{Bolintineanu}, \binits{D.S.}},
\bauthor{\bsnm{Grest}, \binits{G.S.}},
\bauthor{\bsnm{Lechman}, \binits{J.B.}},
\bauthor{\bsnm{Plimpton}, \binits{S.J.}},
\bauthor{\bsnm{Srivastava}, \binits{I.}},
\bauthor{\bsnm{Silbert}, \binits{L.E.}}:
\batitle{Granular packings with sliding, rolling, and twisting friction}.
\bjtitle{Physical Review E}
\bvolume{102}(\bissue{3}),
\bfpage{032903}
(\byear{2020})
\end{barticle}
\endbibitem

\bibitem[\protect\citeauthoryear{Guy et~al.}{2015}]{Guy_2015}
\begin{barticle}
\bauthor{\bsnm{Guy}, \binits{B.M.}},
\bauthor{\bsnm{Hermes}, \binits{M.}},
\bauthor{\bsnm{Poon}, \binits{W.C.K.}}:
\batitle{Towards a unified description of the rheology of hard-particle
  suspensions}.
\bjtitle{Phys. Rev. Lett.}
\bvolume{115},
\bfpage{088304}
(\byear{2015})
\end{barticle}
\endbibitem

\bibitem[\protect\citeauthoryear{Singh et~al.}{2018}]{Singh_2018}
\begin{barticle}
\bauthor{\bsnm{Singh}, \binits{A.}},
\bauthor{\bsnm{Mari}, \binits{R.}},
\bauthor{\bsnm{Denn}, \binits{M.M.}},
\bauthor{\bsnm{Morris}, \binits{J.F.}}:
\batitle{A constitutive model for simple shear of dense frictional
  suspensions}.
\bjtitle{J. Rheol.}
\bvolume{62}(\bissue{2}),
\bfpage{457}--\blpage{468}
(\byear{2018})
\end{barticle}
\endbibitem

\bibitem[\protect\citeauthoryear{More and Ardekani}{2020}]{More_2020}
\begin{barticle}
\bauthor{\bsnm{More}, \binits{R.}},
\bauthor{\bsnm{Ardekani}, \binits{A.}}:
\batitle{Roughness induced shear thickening in frictional non-brownian
  suspensions: A numerical study}.
\bjtitle{Journal of Rheology}
\bvolume{64}(\bissue{2}),
\bfpage{283}--\blpage{297}
(\byear{2020})
\end{barticle}
\endbibitem

\bibitem[\protect\citeauthoryear{Pradeep et~al.}{2021}]{Pradeep_2021}
\begin{barticle}
\bauthor{\bsnm{Pradeep}, \binits{S.}},
\bauthor{\bsnm{Nabizadeh}, \binits{M.}},
\bauthor{\bsnm{Jacob}, \binits{A.R.}},
\bauthor{\bsnm{Jamali}, \binits{S.}},
\bauthor{\bsnm{Hsiao}, \binits{L.C.}}:
\batitle{Jamming distance dictates colloidal shear thickening}.
\bjtitle{Physical Review Letters}
\bvolume{127}(\bissue{15}),
\bfpage{158002}
(\byear{2021})
\end{barticle}
\endbibitem

\bibitem[\protect\citeauthoryear{Maranzano and Wagner}{2001}]{Maranzano_2001}
\begin{barticle}
\bauthor{\bsnm{Maranzano}, \binits{B.J.}},
\bauthor{\bsnm{Wagner}, \binits{N.J.}}:
\batitle{The effects of particle size on reversible shear thickening of
  concentrated colloidal dispersions}.
\bjtitle{J. Chem. Phys.}
\bvolume{114},
\bfpage{10514}--\blpage{527}
(\byear{2001})
\end{barticle}
\endbibitem

\bibitem[\protect\citeauthoryear{Gopalakrishnan and
  Zukoski}{2004}]{Gopalakrishnan_2004}
\begin{barticle}
\bauthor{\bsnm{Gopalakrishnan}, \binits{V.}},
\bauthor{\bsnm{Zukoski}, \binits{C.}}:
\batitle{Effect of attractions on shear thickening in dense suspensions}.
\bjtitle{J. Rheol.}
\bvolume{48}(\bissue{6}),
\bfpage{1321}--\blpage{1344}
(\byear{2004})
\end{barticle}
\endbibitem

\bibitem[\protect\citeauthoryear{Brown et~al.}{2010}]{Brown_2010}
\begin{barticle}
\bauthor{\bsnm{Brown}, \binits{E.}},
\bauthor{\bsnm{Forman}, \binits{N.A.}},
\bauthor{\bsnm{Orellana}, \binits{C.S.}},
\bauthor{\bsnm{Hanjun}, \binits{Z.}},
\bauthor{\bsnm{Maynor}, \binits{B.W.}},
\bauthor{\bsnm{Betts}, \binits{D.E.}},
\bauthor{\bsnm{DeSimone}, \binits{J.M.}},
\bauthor{\bsnm{Jaeger}, \binits{H.M.}}:
\batitle{Generality of shear thickening in dense suspensions}.
\bjtitle{Nat. Mater.}
\bvolume{9},
\bfpage{220}--\blpage{224}
(\byear{2010})
\end{barticle}
\endbibitem

\bibitem[\protect\citeauthoryear{Larson}{1999}]{Larson_1999}
\begin{bbook}
\bauthor{\bsnm{Larson}, \binits{R.G.}}:
\bbtitle{The Structure and Rheology of Complex Fluids}.
\bpublisher{Oxford University Press},
\blocation{New York \& Oxford}
(\byear{1999})
\end{bbook}
\endbibitem

\bibitem[\protect\citeauthoryear{Zaccarelli}{2007}]{Zaccarelli_2007}
\begin{barticle}
\bauthor{\bsnm{Zaccarelli}, \binits{E.}}:
\batitle{Colloidal gels: equilibrium and non-equilibrium routes}.
\bjtitle{Journal of Physics: Condensed Matter}
\bvolume{19}(\bissue{32}),
\bfpage{323101}
(\byear{2007})
\end{barticle}
\endbibitem

\bibitem[\protect\citeauthoryear{Richards et~al.}{2020}]{Richards_2020}
\begin{barticle}
\bauthor{\bsnm{Richards}, \binits{J.A.}},
\bauthor{\bsnm{Guy}, \binits{B.M.}},
\bauthor{\bsnm{Blanco}, \binits{E.}},
\bauthor{\bsnm{Hermes}, \binits{M.}},
\bauthor{\bsnm{Poy}, \binits{G.}},
\bauthor{\bsnm{Poon}, \binits{W.C.}}:
\batitle{The role of friction in the yielding of adhesive non-brownian
  suspensions}.
\bjtitle{Journal of Rheology}
\bvolume{64}(\bissue{2}),
\bfpage{405}--\blpage{412}
(\byear{2020})
\end{barticle}
\endbibitem

\bibitem[\protect\citeauthoryear{Cates et~al.}{1998}]{Cates_1998a}
\begin{barticle}
\bauthor{\bsnm{Cates}, \binits{M.E.}},
\bauthor{\bsnm{Wittmer}, \binits{J.P.}},
\bauthor{\bsnm{Bouchaud}, \binits{J.-P.}},
\bauthor{\bsnm{Claudin}, \binits{P.}}:
\batitle{Jamming, force chains, and fragile matter}.
\bjtitle{Phys. Rev. Lett.}
\bvolume{81},
\bfpage{1841}--\blpage{1844}
(\byear{1998})
\end{barticle}
\endbibitem

\bibitem[\protect\citeauthoryear{Radjai et~al.}{1998}]{Radjai_1998}
\begin{barticle}
\bauthor{\bsnm{Radjai}, \binits{F.}},
\bauthor{\bsnm{Wolf}, \binits{D.E.}},
\bauthor{\bsnm{Jean}, \binits{M.}},
\bauthor{\bsnm{Moreau}, \binits{J.-J.}}:
\batitle{Bimodal character of stress transmission in granular packings}.
\bjtitle{Phys. Rev. Lett.}
\bvolume{80}(\bissue{1}),
\bfpage{61}
(\byear{1998})
\end{barticle}
\endbibitem

\bibitem[\protect\citeauthoryear{Sarkar et~al.}{2013}]{Sarkar_2013}
\begin{barticle}
\bauthor{\bsnm{Sarkar}, \binits{S.}},
\bauthor{\bsnm{Bi}, \binits{D.}},
\bauthor{\bsnm{Zhang}, \binits{J.}},
\bauthor{\bsnm{Behringer}, \binits{R.}},
\bauthor{\bsnm{Chakraborty}, \binits{B.}}:
\batitle{Origin of rigidity in dry granular solids}.
\bjtitle{Physical review letters}
\bvolume{111}(\bissue{6}),
\bfpage{068301}
(\byear{2013})
\end{barticle}
\endbibitem

\bibitem[\protect\citeauthoryear{Sarkar and Chakraborty}{2015}]{Sarkar_2015}
\begin{barticle}
\bauthor{\bsnm{Sarkar}, \binits{S.}},
\bauthor{\bsnm{Chakraborty}, \binits{B.}}:
\batitle{Shear-induced rigidity in athermal materials: a unified statistical
  framework}.
\bjtitle{Physical Review E}
\bvolume{91}(\bissue{4}),
\bfpage{042201}
(\byear{2015})
\end{barticle}
\endbibitem

\bibitem[\protect\citeauthoryear{Sarkar et~al.}{2016}]{Sarkar_2016}
\begin{barticle}
\bauthor{\bsnm{Sarkar}, \binits{S.}},
\bauthor{\bsnm{Bi}, \binits{D.}},
\bauthor{\bsnm{Zhang}, \binits{J.}},
\bauthor{\bsnm{Ren}, \binits{J.}},
\bauthor{\bsnm{Behringer}, \binits{R.P.}},
\bauthor{\bsnm{Chakraborty}, \binits{B.}}:
\batitle{Shear-induced rigidity of frictional particles: Analysis of emergent
  order in stress space}.
\bjtitle{Physical review E}
\bvolume{93}(\bissue{4}),
\bfpage{042901}
(\byear{2016})
\end{barticle}
\endbibitem

\bibitem[\protect\citeauthoryear{Thomas et~al.}{2018}]{thomas2018microscopic}
\begin{barticle}
\bauthor{\bsnm{Thomas}, \binits{J.E.}},
\bauthor{\bsnm{Ramola}, \binits{K.}},
\bauthor{\bsnm{Singh}, \binits{A.}},
\bauthor{\bsnm{Mari}, \binits{R.}},
\bauthor{\bsnm{Morris}, \binits{J.F.}},
\bauthor{\bsnm{Chakraborty}, \binits{B.}}:
\batitle{Microscopic origin of frictional rheology in dense suspensions:
  correlations in force space}.
\bjtitle{Phys. Rev. Lett.}
\bvolume{121}(\bissue{12}),
\bfpage{128002}
(\byear{2018})
\end{barticle}
\endbibitem

\bibitem[\protect\citeauthoryear{Royer et~al.}{2016}]{Royer_2016}
\begin{barticle}
\bauthor{\bsnm{Royer}, \binits{J.R.}},
\bauthor{\bsnm{Blair}, \binits{D.L.}},
\bauthor{\bsnm{Hudson}, \binits{S.D.}}:
\batitle{Rheological signature of frictional interactions in shear thickening
  suspensions}.
\bjtitle{Phys. Rev. Lett.}
\bvolume{116}(\bissue{18}),
\bfpage{188301}
(\byear{2016})
\end{barticle}
\endbibitem

\bibitem[\protect\citeauthoryear{Jamali et~al.}{2020}]{Jamali_2020}
\begin{barticle}
\bauthor{\bsnm{Jamali}, \binits{S.}},
\bauthor{\bsnm{Del~Gado}, \binits{E.}},
\bauthor{\bsnm{Morris}, \binits{J.F.}}:
\batitle{Rheology discussions: The physics of dense suspensions}.
\bjtitle{Journal of Rheology}
\bvolume{64}(\bissue{6}),
\bfpage{1501}--\blpage{1524}
(\byear{2020})
\end{barticle}
\endbibitem

\bibitem[\protect\citeauthoryear{Lee et~al.}{2020}]{Lee_2020}
\begin{barticle}
\bauthor{\bsnm{Lee}, \binits{Y.-F.}},
\bauthor{\bsnm{Luo}, \binits{Y.}},
\bauthor{\bsnm{Brown}, \binits{S.C.}},
\bauthor{\bsnm{Wagner}, \binits{N.J.}}:
\batitle{Experimental test of a frictional contact model for shear thickening
  in concentrated colloidal suspensions}.
\bjtitle{Journal of Rheology}
\bvolume{64}(\bissue{2}),
\bfpage{267}--\blpage{282}
(\byear{2020})
\end{barticle}
\endbibitem

\bibitem[\protect\citeauthoryear{Boromand et~al.}{2018}]{Boromand_2018}
\begin{barticle}
\bauthor{\bsnm{Boromand}, \binits{A.}},
\bauthor{\bsnm{Jamali}, \binits{S.}},
\bauthor{\bsnm{Grove}, \binits{B.}},
\bauthor{\bsnm{Maia}, \binits{J.M.}}:
\batitle{A generalized frictional and hydrodynamic model of the dynamics and
  structure of dense colloidal suspensions}.
\bjtitle{Journal of Rheology}
\bvolume{62}(\bissue{4}),
\bfpage{905}--\blpage{918}
(\byear{2018})
\end{barticle}
\endbibitem

\bibitem[\protect\citeauthoryear{Gameiro et~al.}{2020}]{Gameiro_2020}
\begin{barticle}
\bauthor{\bsnm{Gameiro}, \binits{M.}},
\bauthor{\bsnm{Singh}, \binits{A.}},
\bauthor{\bsnm{Kondic}, \binits{L.}},
\bauthor{\bsnm{Mischaikow}, \binits{K.}},
\bauthor{\bsnm{Morris}, \binits{J.F.}}:
\batitle{Interaction network analysis in shear thickening suspensions}.
\bjtitle{Physical Review Fluids}
\bvolume{5}(\bissue{3}),
\bfpage{034307}
(\byear{2020})
\end{barticle}
\endbibitem

\bibitem[\protect\citeauthoryear{Edens et~al.}{2021}]{Edens_2021}
\begin{barticle}
\bauthor{\bsnm{Edens}, \binits{L.E.}},
\bauthor{\bsnm{Alvarado}, \binits{E.G.}},
\bauthor{\bsnm{Singh}, \binits{A.}},
\bauthor{\bsnm{Morris}, \binits{J.F.}},
\bauthor{\bsnm{Schenter}, \binits{G.K.}},
\bauthor{\bsnm{Chun}, \binits{J.}},
\bauthor{\bsnm{Clark}, \binits{A.E.}}:
\batitle{Shear stress dependence of force networks in 3d dense suspensions}.
\bjtitle{Soft Matter}
\bvolume{17}(\bissue{32}),
\bfpage{7476}--\blpage{7486}
(\byear{2021})
\end{barticle}
\endbibitem

\bibitem[\protect\citeauthoryear{Singh et~al.}{2019}]{Singh_2019}
\begin{barticle}
\bauthor{\bsnm{Singh}, \binits{A.}},
\bauthor{\bsnm{Pednekar}, \binits{S.}},
\bauthor{\bsnm{Chun}, \binits{J.}},
\bauthor{\bsnm{Denn}, \binits{M.M.}},
\bauthor{\bsnm{Morris}, \binits{J.F.}}:
\batitle{From yielding to shear jamming in a cohesive frictional suspension}.
\bjtitle{Phys. Rev. Lett.}
\bvolume{122}(\bissue{9}),
\bfpage{098004}
(\byear{2019})
\end{barticle}
\endbibitem

\bibitem[\protect\citeauthoryear{Nabizadeh et~al.}{2022}]{Nabizadeh_2022}
\begin{barticle}
\bauthor{\bsnm{Nabizadeh}, \binits{M.}},
\bauthor{\bsnm{Singh}, \binits{A.}},
\bauthor{\bsnm{Jamali}, \binits{S.}}:
\batitle{Structure and dynamics of force clusters and networks in shear
  thickening suspensions}.
\bjtitle{Physical Review Letters}
\bvolume{129}(\bissue{6}),
\bfpage{068001}
(\byear{2022})
\end{barticle}
\endbibitem

\bibitem[\protect\citeauthoryear{van~der Naald et~al.}{2023}]{Naald_2023}
\begin{botherref}
\oauthor{\bsnm{Naald}, \binits{M.}},
\oauthor{\bsnm{Singh}, \binits{A.}},
\oauthor{\bsnm{Eid}, \binits{T.}},
\oauthor{\bsnm{Tang}, \binits{K.}},
\oauthor{\bsnm{Pablo}, \binits{J.}},
\oauthor{\bsnm{Jaeger}, \binits{H.}}:
Minimally rigid clusters in dense suspension flow
(2023)
\end{botherref}
\endbibitem

\bibitem[\protect\citeauthoryear{Goyal et~al.}{2022}]{Goyal_2022}
\begin{botherref}
\oauthor{\bsnm{Goyal}, \binits{A.}},
\oauthor{\bsnm{Martys}, \binits{N.S.}},
\oauthor{\bsnm{Del~Gado}, \binits{E.}}:
Flow induced rigidity percolation in shear thickening suspensions.
arXiv preprint arXiv:2210.00337
(2022)
\end{botherref}
\endbibitem

\bibitem[\protect\citeauthoryear{Comtet et~al.}{2017}]{Comtet_2017}
\begin{barticle}
\bauthor{\bsnm{Comtet}, \binits{J.}},
\bauthor{\bsnm{Chatt{\'e}}, \binits{G.}},
\bauthor{\bsnm{Nigu{\`e}s}, \binits{A.}},
\bauthor{\bsnm{Bocquet}, \binits{L.}},
\bauthor{\bsnm{Siria}, \binits{A.}},
\bauthor{\bsnm{Colin}, \binits{A.}}:
\batitle{Pairwise frictional profile between particles determines discontinuous
  shear thickening transition in non-colloidal suspensions.}
\bjtitle{Nat. Comm.}
\bvolume{8},
\bfpage{15633}
(\byear{2017})
\end{barticle}
\endbibitem

\bibitem[\protect\citeauthoryear{Singh}{2022}]{Singh_2022a}
\begin{barticle}
\bauthor{\bsnm{Singh}, \binits{A.}}:
\batitle{Shear thickening in dense suspension: A master-curve and “roll” of
  friction}.
\bjtitle{Science Talks}
\bvolume{3},
\bfpage{100028}
(\byear{2022})
\end{barticle}
\endbibitem

\bibitem[\protect\citeauthoryear{Castle et~al.}{1996}]{Castle_1996}
\begin{barticle}
\bauthor{\bsnm{Castle}, \binits{J.}},
\bauthor{\bsnm{Farid}, \binits{A.}},
\bauthor{\bsnm{Woodcock}, \binits{L.V.}}:
\batitle{The effect of surface friction on the rheology of hard-sphere
  colloids}.
\bjtitle{Progr. Colloid Polym. Sci.}
\bvolume{100},
\bfpage{259}--\blpage{265}
(\byear{1996})
\end{barticle}
\endbibitem

\bibitem[\protect\citeauthoryear{Lootens et~al.}{2003}]{Lootens_2003}
\begin{barticle}
\bauthor{\bsnm{Lootens}, \binits{D.}},
\bauthor{\bsnm{Van~Damme}, \binits{H.}},
\bauthor{\bsnm{H\'ebraud}, \binits{P.}}:
\batitle{Giant stress fluctuations at the jamming transition}.
\bjtitle{Phys. Rev. Lett.}
\bvolume{90},
\bfpage{178301}
(\byear{2003})
\end{barticle}
\endbibitem

\bibitem[\protect\citeauthoryear{Lootens et~al.}{2004}]{Lootens_2004}
\begin{barticle}
\bauthor{\bsnm{Lootens}, \binits{D.}},
\bauthor{\bsnm{H\'ebraud}, \binits{P.}},
\bauthor{\bsnm{L\'ecolier}, \binits{E.}},
\bauthor{\bsnm{Van~Damme}, \binits{H.}}:
\batitle{Gelation, shear-thinning and shear-thickening in cement slurries}.
\bjtitle{Oil Gas Sci. Technol.- Rev. IFP.}
\bvolume{59}(\bissue{1}),
\bfpage{31}--\blpage{40}
(\byear{2004})
\end{barticle}
\endbibitem

\bibitem[\protect\citeauthoryear{Hsiao et~al.}{2017}]{Hsiao_2017}
\begin{barticle}
\bauthor{\bsnm{Hsiao}, \binits{L.C.}},
\bauthor{\bsnm{Jamali}, \binits{S.}},
\bauthor{\bsnm{Glynos}, \binits{E.}},
\bauthor{\bsnm{Green}, \binits{P.F.}},
\bauthor{\bsnm{Larson}, \binits{R.G.}},
\bauthor{\bsnm{Solomon}, \binits{M.J.}}:
\batitle{Rheological state diagrams for rough colloids in shear flow}.
\bjtitle{Phys. Rev. Lett.}
\bvolume{119}(\bissue{15}),
\bfpage{158001}
(\byear{2017})
\end{barticle}
\endbibitem

\bibitem[\protect\citeauthoryear{James et~al.}{2018}]{james2018interparticle}
\begin{barticle}
\bauthor{\bsnm{James}, \binits{N.M.}},
\bauthor{\bsnm{Han}, \binits{E.}},
\bauthor{\bsnm{Cruz}, \binits{R.A.L.}},
\bauthor{\bsnm{Jureller}, \binits{J.}},
\bauthor{\bsnm{Jaeger}, \binits{H.M.}}:
\batitle{Interparticle hydrogen bonding can elicit shear jamming in dense
  suspensions}.
\bjtitle{Nat. Mater.}
\bvolume{17}(\bissue{11}),
\bfpage{965}
(\byear{2018})
\end{barticle}
\endbibitem

\bibitem[\protect\citeauthoryear{James et~al.}{2019}]{James_2019}
\begin{barticle}
\bauthor{\bsnm{James}, \binits{N.M.}},
\bauthor{\bsnm{Hsu}, \binits{C.-P.}},
\bauthor{\bsnm{Spencer}, \binits{N.D.}},
\bauthor{\bsnm{Jaeger}, \binits{H.M.}},
\bauthor{\bsnm{Isa}, \binits{L.}}:
\batitle{Tuning interparticle hydrogen bonding in shear-jamming suspensions:
  Kinetic effects and consequences for tribology and rheology}.
\bjtitle{J. Phys. Chem. Lett.}
\bvolume{10}(\bissue{8}),
\bfpage{1663}--\blpage{1668}
(\byear{2019})
\end{barticle}
\endbibitem

\bibitem[\protect\citeauthoryear{Laun}{1984}]{Laun_1984}
\begin{barticle}
\bauthor{\bsnm{Laun}, \binits{H.M.}}:
\batitle{Rheological properties of aqueous polymer dispersions}.
\bjtitle{Angew. Makromol. Chem.}
\bvolume{123}(\bissue{1}),
\bfpage{335}--\blpage{359}
(\byear{1984})
\end{barticle}
\endbibitem

\bibitem[\protect\citeauthoryear{Xu et~al.}{2020}]{Xu_2020}
\begin{barticle}
\bauthor{\bsnm{Xu}, \binits{Q.}},
\bauthor{\bsnm{Singh}, \binits{A.}},
\bauthor{\bsnm{Jaeger}, \binits{H.M.}}:
\batitle{Stress fluctuations and shear thickening in dense granular
  suspensions}.
\bjtitle{Journal of Rheology}
\bvolume{64}(\bissue{2}),
\bfpage{321}--\blpage{328}
(\byear{2020})
\end{barticle}
\endbibitem

\bibitem[\protect\citeauthoryear{van~der Naald et~al.}{2021}]{Naald_2021}
\begin{botherref}
\oauthor{\bsnm{Naald}, \binits{M.}},
\oauthor{\bsnm{Zhao}, \binits{L.}},
\oauthor{\bsnm{Jackson}, \binits{G.L.}},
\oauthor{\bsnm{Jaeger}, \binits{H.M.}}:
The role of solvent molecular weight in shear thickening and shear jamming.
Soft Matter
(2021)
\end{botherref}
\endbibitem

\bibitem[\protect\citeauthoryear{Chen et~al.}{2023}]{Chen_2022}
\begin{barticle}
\bauthor{\bsnm{Chen}, \binits{C.}},
\bauthor{\bsnm{Naald}, \binits{M.}},
\bauthor{\bsnm{Singh}, \binits{A.}},
\bauthor{\bsnm{Dolinski}, \binits{N.D.}},
\bauthor{\bsnm{Jackson}, \binits{G.L.}},
\bauthor{\bsnm{Jaeger}, \binits{H.M.}},
\bauthor{\bsnm{Rowan}, \binits{S.J.}},
\bauthor{\bsnm{Pablo}, \binits{J.J.}}:
\batitle{Leveraging the polymer glass transition to access thermally switchable
  shear jamming suspensions}.
\bjtitle{ACS Central Science}
\bvolume{9}(\bissue{4}),
\bfpage{639}--\blpage{647}
(\byear{2023})
\end{barticle}
\endbibitem

\end{thebibliography}

\end{document}